# A low-frequency study of recently identified double-double radio galaxies

S. Nandi[1,2] *, D.J. Saikia[3,4], R. Roy[3], P. Dabhade[3], Y. Wadadekar[4], J. Larsson[2], M. Baes[5], H.C. Chandola[6] and M. Singh[7]

[1] *Indian Institute of Astrophysics, Bangalore 560034, India*
[2] *KTH, Department of Physics, and the Oskar Klein Centre, AlbaNova, SE-106 91 Stockholm, Sweden*
[3] *Inter-University Centre for Astronomy and Astrophysics (IUCAA), Post Bag 4, Ganeshkhind, Pune 411007, India*
[4] *National Centre for Radio Astrophysics, TIFR, Post Bag 3, Ganeshkhind, Pune 411 007, India*
[5] *Sterrenkundig Observatorium, Universiteit Gent, Krijgslaan 281 S9, B-9000 Gent, Belgium*
[6] *Department of Physics, Kumaun University, Nainital 263 001, India*
[7] *Aryabhatta Research Institute of Observational Sciences (ARIES), Manora Peak, Nainital, 263 129, India*

**ABSTRACT**
In order to understand the possible mechanisms of recurrent jet activity in radio galaxies and quasars, which are still unclear, we have identified such sources with a large range of linear sizes (220 − 917 kpc), and hence time scales of episodic activity. Here we present high-sensitivity 607-MHz Giant Metrewave Radio Telescope (GMRT) images of 21 possible double-double radio galaxies (DDRGs) identified from the FIRST survey to confirm their episodic nature. These GMRT observations show that none of the inner compact components suspected to be hot-spots of the inner doubles are cores having a flat radio spectrum, confirming the episodic nature of these radio sources. We have indentified a new DDRG with a candidate quasar, and have estimated the upper spectral age limits for eight sources which showed marginal evidence of steepening at higher frequencies. The estimated age limits (11 − 52 Myr) are smaller than those of the large-sized ($\sim$ 1 Mpc) DDRGs.

## 1 INTRODUCTION

Double-double radio galaxies (DDRGs) can be characterized by the presence of a second pair, and occasionally a third pair, of radio lobes driven by the same central active galactic nucleus (AGN). Such radio sources are relatively rare examples of AGNs that undergo multiple cycles of jet activity. In most cases, the diffuse outer double lobes appear reasonably well aligned with the inner ones and extend from $\sim 10^2$ kpc up to a few Mpc. The bright inner doubles span from ∼10 pc to several 100 kpc (Saikia & Jamrozy 2009). The mechanisms for the interruption of these bipolar relativistic jet flows, the effects of and on their ambient medium, and the timescales of their duty cycle have been the subject of a number of investigations (e.g., Jamrozy et al. 2008; Konar et al. 2012; Mahatma et al. 2018; Brienza et al. 2018). A large sample of DDRGs with a wide range of sizes is required to address these questions satisfactorily.

In order to increase the number of known DDRGs, Nandi & Saikia (2012) made a visual inspection of the radio structures and optical identifications of each of the 242 sources listed by Proctor (2011) as candidate DDRGs. However, we could not confirm many of the classifications. For example, in a substantial number of cases, a radio core and a prominent knot or component resembled an inner double and was classified as a DDRG. From a detailed examination of the FIRST (Faint Images of the Radio Sky at Twenty centimeters; Becker et al. 1995) images and optical data, only 23 of these were found likely to be actual DDRGs, with diffuse outer lobes and more compact inner double-lobed structures. Their optical hosts are co-incident with the radio cores or located between the inner compact components. Though early studies indicated that repeated jet activity is not very common among radio sources smaller than ∼1 Mpc (Schoenmakers et al. 2001) and large linear sizes are charac-

* e-mail: sumana.nandi@iiap.res.in

Table 1. The sample of galaxies observed with the GMRT at 607 MHz.

| Name (1) | Opt. Id. (2) | RA hh:mm:ss.ss (3) | Dec dd:mm:ss.ss (4) | $z$ (5) | Obs. date (6) | Flux cal used (7) | Phase cal used (8) | Beam size ($'' \times ''$) (9) | BPA ($°$) (10) | T (hr) |
|---|---|---|---|---|---|---|---|---|---|---|
| J0746+4526 | Q | 07:46:17.92 | +45:26:34.47 | 0.550 | 28-11-2012 | 3C147 | 0713+438 | 6.83×4.71 | 73 | 3.6 |
| J0804+5809 | Q | 08:04:42.79 | +58:09:34.94 | (0.300) | 28-11-2012 | 3C147, 3C286 | 0614+607 | 6.08×4.82 | 330 | 3.6 |
| J0855+4204 | G | 08:55:49.15 | +42:04:20.12 | (0.238) | 01-12-2012 | 3C147 | 0741+312 | 7.09×4.76 | 79 | 3.6 |
| J0910+0345 | G | 09:10:59.10 | +03:45:31.68 | (0.588) | 07-02-2013 | 3C48, 3C286 | 0842+185 0943−083 | 5.52×4.73 | 221 | 3.6 |
| J1039+0536 | G | 10:39:28.21 | +05:36:13.61 | 0.0908 | 29-12-2012 | 3C147, 3C286 | 1120+143 | 4.44×4.06 | 84 | 3.6 |
| J1103+0636 | G | 11:03:13.29 | +06:36:16.00 | 0.4405 | 01-12-2012 | 3C147, 3C286 | 1120+143 | 4.81×4.80 | 40 | 3.6 |
| J1208+0821 | G | 12:08:56.78 | +08:21:38.57 | 0.5841 | 25-06-2014 | 3C286 | 1150-003 | 5.26×4.71 | 274 | 2.3 |
| J1238+1602 |  | 12:38:21.20 | +16:02:41.43 |  | 24-03-2013 | 3C147, 3C286 | 1330+251 | 5.58×3.31 | 65 | 3.0 |
| J1240+2122 | G | 12:40:13.48 | +21:22:33.04 | (0.634) | 03-12-2012 | 3C147, 3C286 | 1254+116 | 4.77×3.79 | 61 | 3.6 |
| J1326+1924 | G | 13:26:13.67 | +19:24:23.75 | 0.1762 | 25-06-2014 | 3C286 | 1330+251 | 7.73×4.24 | 83 | 1.6 |
| J1328+2752 | G | 13:28:48.45 | +27:52:27.81 | 0.0911 | 24-03-2013 | 3C286 | 1330+251 | 5.27×4.22 | 69 | 3.6 |
| J1344−0030 | G | 13:44:46.92 | −00:30:09.28 | 0.5800 | 26-06-2014 | 3C286 | 1354-021 | 5.08×3.82 | 81 | 2.6 |
| J1407+5132 | G | 14:07:18.49 | +51:32:04.88 | 0.3404 | 26-06-2014 | 3C286 | 1400+621 | 7.27×3.73 | 279 | 1.3 |
| J1500+1542 | G | 15:00:55.18 | +15:42:40.64 | (0.456) | 28-06-2014 | 3C286 | 1445+099 | 4.46×4.37 | 82 | 2.0 |
| J1521+5214 | G | 15:21:05.90 | +52:14:40.15 | (0.670) | 28-06-2014 | 3C286, 3C48 | 1438+621 | 7.61×4.60 | 291 | 2.2 |
| J1538−0242 | G | 15:38:41.31 | −02:42:05.52 | (0.598) | 28-06-2014 | 3C286, 3C48 | 1438+621 | 4.71×3.68 | 52 | 2.3 |
| J1545+5047 | G | 15:45:17.21 | +50:47:54.18 | 0.4309 | 23-06-2014 | 3C286, 3C48 | 1634+627 | 6.41×3.62 | 278 | 2.6 |
| J1605+0711 | G | 16:05:13.74 | +07:11:52.56 | 0.311 | 16-07-2014 | 3C286, 3C48 | 1557-000 | 5.97×3.93 | 68 | 2.0 |
| J1627+2906 | G | 16:27:54.63 | +29:06:20.00 | (0.722) | 27-06-2014 | 3C286, 3C48 | 1609+266 | 7.25×4.67 | 271 | 2.3 |
| J1649+4133 |  | 16:49:28.32 | +41:33:41.58 |  | 25-06-2014 | 3C286, 3C48 | 1635+381 | 6.95×3.96 | 286 | 2.6 |
| J1705+3940 | G | 17:05:17.83 | +39:40:29.25 | (0.778) | 26-06-2014 | 3C286, 3C48 | 1613+342 | 6.88×4.40 | 288 | 1.6 |

Column 1: source name; Column 2: optical identification; G and Q represent galaxy and quasar respectively while for two sources which are faint and whose identifications are uncertain the entries have been left blank; Columns 3 and 4: right ascension and declination of the optical objects in J2000 co-ordinates; Column 5: redshift. The photometric redshifts are enclosed in parentheses. Column 6: the dates of GMRT observations; Columns 7 and 8: the names of the calibrators used for each observation; Columns 9 and 10: the major and minor axes of the restoring beam in arcsec and beam position angle (BPA) in degrees; Column 11: observing time on source in units of hr.

teristic of most of the known DDRGs, this did not appear to be the case for the DDRGs identified from the FIRST survey. Certainly, improved statistics on large- and small-sized DDRGs will be helpful to analyze how recurrent activity influences their evolution process, and help constrain models of recurrent activity. From the FIRST images at 1400 MHz Nandi & Saikia (2012) found that the median sizes of the inner and outer doubles of these DDRGs are (∼75 and 530 kpc respectively) comparable to the four smaller-sized radio sources like 3C293, 3C219, 4C02.27 and Cyg A that show intermittent jet activity (Saikia & Jamrozy 2009).

In addition, the spectral ageing analysis of 3C293 showed that the interruption time between the two activity epochs is only $10^5$yr (Joshi et al. 2011), which is significantly less than the other large-sized DDRGs with typical interruption timescale $10^7$yr to $10^8$yr. Therefore, evidence for the existence of such smaller-sized DDRGs suggests that a wide range of time scales of episodic jet cycles in AGNs is possible. Kuźmicz et al. (2017) investigated the optical host properties of 74 restarting radio sources which includes our sample as well. Their results show that the black hole mass for these restarting sources are comparable to those of classical FRII radio galaxies, while the concentration indices (CI; the ratio of radius containing 90% of Petrosian flux to the radius containing 50% of Petrosian flux at $r$ band) are significantly smaller in case of episodic sources. This suggests that the hosts of DDRGs are often associated with past or ongoing galaxy interactions. They also found that the hosts of restarting radio sources tend to contain more young stars than the hosts of typical radio sources. Usually for late type galaxies the CI value is less than 2.86 while for early type galaxies CI value is greater than 2.86 (Nakamura et al. 2003).

We started 607-MHz GMRT observation of these candidate DDRGs to image the different components, estimate the flux densities and spectral indices of the outer and inner lobes in order to confirm their episodic nature. From our list we have excluded two sources, J1158+2621 and J1706+4340 because they have been already investigated using GMRT by Konar et al. (2013) and Marecki et al. (2016) respectively. In this paper we present the GMRT observations of the remaining 21 sources (Table 1). We observed these candidates in cycles 23 and 26 under proposal codes 23_056 & 26_030. The detailed study of J0746+4526 and J1328+2752 from this sample have been published in Nandi et al. (2014) and Nandi et al. (2017). Throughout the paper we assume a cosmology with $H_o$=71 km s$^{-1}$ Mpc$^{-1}$, $\Omega_{\rm m}$=0.27 and $\Omega_{\rm vac}$=0.73. The observations and radio data reduction procedures are described in Section 2. In Section 3, we present the results from our GMRT observations. This Section also includes a description of each individual source. A discussion of spectral ages is presented in Section 4, while the results are summarised in Section 5.

## 2 OBSERVATIONS AND DATA REDUCTION

The details of the GMRT interferometric observations of our sample are given in Table 1. The observations were made following the usual protocol of observing flux density and phase calibrators with the observations of the target source. The observing bandwidth is 32 MHz. At the beginning and at the end of each observing run, one of the flux density calibrators 3C48, 3C286 or 3C147 was observed for about 15 min. All flux densities are on the Perley & Butler (2013) scale using the latest VLA values. The phase calibrators were observed for 5 min after each of several 20 min exposures of the target sources. The data reduction was performed using the NRAO Astronomical Image Processing System (AIPS)[1]. After removing bad antennas and the strong radio frequency interference (RFI), standard flux density and phase calibration were applied to the sources. Around 20% data were edited out. We use both RR and LL polarization data for imaging. The total field of view of each source was split into 25 facets. This process helps to keep each small facet as a plane surface. Several rounds of self calibration were performed to produce the best possible images. All final images were corrected for the primary beam pattern of the GMRT. For some sources which either had artefacts and/or higher than expected rms values, and also for purposes of comparison we used the SPAM pipeline (Intema et al. 2017). Processing of data using the SPAM pipeline was carried out using Version 17.9.22 of the software with all default settings. The SPAM pipeline applies flux corrections based on measurements of the sky and system temperatures using the prescription described in Intema et al. (2017). SPAM images are made from direction dependent calibration of the data. The best GMRT 607-MHz images which recovered almost the entire integrated flux densities of the sources are presented here. The 1400 MHz images are from the FIRST survey.

## 3 DISCUSSION ON RADIO MORPHOLOGY

The GMRT full-resolution 607-MHz images for all 21 galaxies are presented in the Figures 1 to 4. The optical positions have been marked in each image. We used the Sloan Digital Sky Survey (SDSS) Data Release 14 (DR14)[2] and Panoramic Survey Telescope and Rapid Response System (Pan-STARRS)[3] for optical identification. The observational parameters and the observed properties for both GMRT and FIRST data are presented in Table 2. All FIRST images are available in Nandi & Saikia (2012). To estimate the projected linear size we use GMRT images. For the outer emission without any prominent hotspots we consider the outer-most contours. Here the outer-most contour levels correspond to 3 sigma of image rms noise. Projected linear size for the inner double is the distance between the two bright hotspots. The linear size of each source at 607 MHz is higher than the value noted in Nandi & Saikia (2012). This is because of larger extended emission seen at low frequencies.

---

[1] http://www.aips.nrao.edu software package
[2] https://www.sdss.org/dr14/
[3] https://panstarrs.stsci.edu/

We noticed, their angular sizes increased by ∼10% at 607 MHz. A difference in spectral indices is expected between the outer diffuse lobes which are older and the inner younger components. The spectral index analysis is a useful tool to investigate the second epoch of activity. To estimate the spectral indices $\alpha$ ($S_\nu \propto \nu^{-\alpha}$), the total intensity maps at 607 MHz and 1400 MHz are convolved to a common resolution of ∼5″ to 8″ depending upon each source. From these convolved maps the flux densities at two frequencies have been estimated over similar areas. The errors in the flux densities are approximately 7% at 607 MHz and 5% at 1400 MHz (e.g., Joshi et al. 2011), including calibration errors. The final errors in the spectral indices have been estimated by propagating individual errors in quadrature,

$$\alpha_{\mathrm{err}} = \frac{1}{ln\frac{\nu_1}{\nu_2}} \sqrt{\left(\frac{S1_{\mathrm{err}}}{S1}\right)^2 + \left(\frac{S2_{\mathrm{err}}}{S2}\right)^2}, \qquad (1)$$

Here S1 and S2 are the integrated flux densities at frequencies at $\nu_1$ and $\nu_2$, and $S1_{\mathrm{err}}$ and $S2_{\mathrm{err}}$ are the corresponding flux density errors. The estimated spectral indices are given in Table 3. We note that the angular size of the outer lobes is well above ∼1′ for the sources J0855+4204, J1328+2752, J1407+5132 and J1605+0711. So we may not recover the entire emission of these large sources using their FIRST maps. Thus, we consider the 1.4 GHz flux densities of their outer lobes as lower limits. Hence their spectral indices between 607 MHz and 1.4 GHz actually represent the upper limits. From Table 3 we can see all inner components have relatively steep spectra, demonstrating that these are not core or nuclear components. The variations of spectral indices between the inner and outer doubles are shown in Figure 5. In cases the where the weak ($\lesssim$5 mJy) inner lobes/components are embedded in diffuse structure or the inner lobes are small the errors of corresponding spectral indices can be larger. In Figure 6 we plot the projected linear sizes of the inner and outer lobes versus their spectral indices. The outer doubles tend to have steeper spectra than the inner ones. In the case of J1240+2122 there is a suggestion that the injection spectral indices may be different at the two epochs which requires confirmation from more multi-frequency observations.

We have also calculated the CI values of the host galaxies in our sample for which SDSS (Sloan Digital Sky Survey) and Pan-STARRS data are available. The CI value is higher than 2.86 for the sources J1326+1924 and J1328+2752 (see Table 3). So, according to Nakamura et al. (2003), the hosts of these sources are possibly early type galaxies. In our previuos work (Nandi et al. 2017) we have also shown that J1328+2752 has an elliptical host which is possibly going through primary phase of galaxy evolution. Similarly, SDSS shows the existence of an early type host in J1326+1923 which has an S-shaped (Capetti et al. 2002) radio-morphology. Thus both objects are satisfying the scheme of Nakamura et al. (2003). For 6 sources in our sample (J1103+0636, J1240+212, J1344−0030, J1545+5047, J1627+2906 and J1705+3940), we estimated the upper limits of CI values. Four of which (J1240+212, J1344−0030, J1627+2906 and J1705+3940) have values less than 2.86. For 3 sources in our sample (J0804+5809, J1208+0821 and J1538−0242) the estimated CI values without limits are less than 2.86. Whereas for 5 sources (J0746+4526, J0855+4204, J1039+0536, J1407+5132 and J1605+0711), the estimated

**Table 2.** The observational parameters and observed properties of the sources

| Name | Freq. MHz | rms mJy /b | $S_I$ mJy | Outer comp. | $S_p$ mJy /b | $S_t$ mJy | Inner comp. | $S_p$ mJy /b | $S_t$ mJy | Inner comp. | $S_p$ mJy /b | $S_t$ mJy | Outer comp. | $S_p$ mJy /b | $S_t$ mJy |
|---|---|---|---|---|---|---|---|---|---|---|---|---|---|---|---|
| (1) | (2) | (3) | (4) | (5) | (6) | (7) | (8) | (9) | (10) | (11) | (12) | (13) | (14) | (15) | (16) |
| J0746+4526 | 607 | 0.05 | 465 | NW1 | 3.53 | 93 | NW2 | 17 | 24 | SE2 | 24 | 30 | SE1 | 24 | 319 |
|  | 1400 | 0.16 | 168 | NW1 | 1.11 | 24 | NW2 | 7.91 | 11 | SE2 | 12 | 13 | SE1 | 10 | 119 |
| J0804+5809 | 607 | 0.08 | 556 | Total | 11 | 397 | W2 | 45 | 87 | E2 | 39 | 84 |  |  |  |
|  | 1400 | 0.13 | 159 | W1 | 1.51 | 34 | W2 | 22 | 38 | E2 | 18 | 39 | E1 | 2.03 | 48 |
| J0855+4204† | 607 | 0.07 | 314 | NE1 | 5.15 | 120 | Total | 19 | 46 |  |  |  | SW1 | 5.69 | 143 |
|  | 1400 | 0.10 | 101 | NE1 | 1.80 | 37 | NE2 | 7.33 | 10 | SW2 | 5.97 | 7.10 | SW1 | 2.11 | 43 |
| J0910+0345 | 607 | 0.07 | 223 | W1 | 15 | 59 | Total | 46 | 102 |  |  |  | E1 | 17 | 54 |
|  | 1400 | 0.16 | 81 | W1 | 6.05 | 16 | W2 | 24 | 28 | E2 | 22 | 23 | E1 | 5.70 | 14 |
| J1039+0536 | 608 | 0.06 | 1191 | W1 | 29 | 594 | W2 | 32 | 55 | E2 | 28 | 57 | E1 | 21 | 477 |
|  | 1400 | 0.19 | 519 | W1 | 23 | 252 | W2 | 18 | 29 | E2 | 17 | 29 | E1 | 14 | 203 |
| J1103+0636 | 607 | 0.08 | 229 | W1 | 3.93 | 104 | W2 | 9.74 | 14 | E2 | 11 | 19 | E1 | 3.14 | 91 |
|  | 1400 | 0.13 | 65 | W1 | 1.65 | 25 | W2 | 4.66 | 5.64 | E2 | 5.56 | 6.68 | E1 | 1.50 | 22 |
| J1208+0821 | 607 | 0.07 | 113 | W1 | 4.21 | 51 | W2 | 2.49 | 2.72 | E2 | 2.43 | 3.02 | E1 | 5.48 | 58 |
|  | 1400 | 0.16 | 40 | W1 | 2.17 | 16 | W2 | 1.01 | 1.03 | E2 | 1.11 | 1.12 | E1 | 3.44 | 17 |
| J1238+1602 | 607 | 0.08 | 109 | NW1 | 14 | 48 | NW2 | 5.00 | 8.22 | SE2 | 3.02 | 9.61 | SE1 | 6.65 | 44 |
|  | 1400 | 0.15 | 53 | NW1 | 10 | 25 | NW2 | 3.50 | 4.40 | SE2 | 2.14 | 4.69 | SE1 | 4.65 | 18 |
| J1240+2122 | 607 | 0.20 | 174 | NW1 | 58 | 90 | NW2 | 5.01 | 8.33 | SE2 | 32 | 57 | SE1 | 14 | 19 |
|  | 1400 | 0.14 | 94 | NW1 | 35 | 47 | NW2 | 3.00 | 5.42 | SE2 | 17 | 24 | SE1 | 11 | 13 |
| J1326+1924 | 608 | 0.07 | 154 | W1 | 9.07 | 70 | Total | 8.48 | 17 |  |  |  | E1 | 9.36 | 58 |
|  | 1400 | 0.17 | 47 | W1 | 4.09 | 20 | W2 | 2.55 | 2.85 | E2 | 2.39 | 3.27 | E1 | 4.38 | 15 |
| J1328+2752† | 607 | 0.07 | 427 | NW1 | 3.60 | 164 | NW2 | 20 | 29 | SE2 | 12 | 27 | SE1 | 11 | 200 |
|  | 1400 | 0.14 | 150 | NW1 | 3.17 | 50 | NW2 | 11 | 16 | SE2 | 5.15 | 9.10 | SE1 | 6.67 | 74 |
| J1344−0030 | 607 | 0.06 | 137 | NE1 | 2.05 | 34 | NE2 | 14 | 22 | SW2 | 11 | 15 | SW1 | 3.94 | 64 |
|  | 1400 | 0.14 | 46 | NE1 | 1.58 | 6.69 | NE2 | 8.66 | 12 | SW2 | 7.14 | 8.69 | SW1 | 3.13 | 20 |
| J1407+5132† | 607 | 0.14 | 1277 | NW1 | 23 | 480 | NW2 | 16 | 18 | SE2 | 15 | 16 | SE1 | 31 | 784 |
|  | 1400 | 0.19 | 423 | NW1 | 10 | 159 | NW2 | 5.00 | 5.10 | SE2 | 4.08 | 3.03 | SE1 | 15 | 256 |
| J1500+1542 | 607 | 0.06 | 58 | NW1 | 1.45 | 12 | NW2 | 4.98 | 9.23 | SE2 | 9.57 | 14 | SE1 | 1.71 | 19 |
|  | 1400 | 0.14 | 21 | NW1 | 0.79 | 5.37 | NW2 | 2.64 | 3.91 | SE2 | 5.35 | 6.38 | SE1 | 1.01 | 5.30 |
| J1521+5214 | 607 | 0.08 | 64 | NW1 | 6.23 | 27 | NW2 | 4.53 | 4.94 | SE2 | 8.81 | 9.23 | SE1 | 3.90 | 24 |
|  | 1400 | 0.14 | 25 | NW1 | 1.92 | 7.90 | NW2 | 1.96 | 2.98 | SE2 | 3.72 | 5.72 | SE1 | 1.33 | 6.60 |
| J1538−0242 | 608 | 0.06 | 222 | NW1 | 4.27 | 109 | NW2 | 9.16 | 13 | SE2 | 7.15 | 10 | SE1 | 14 | 95 |
|  | 1400 | 0.13 | 64 | NW1 | 8.02 | 28 | NW2 | 3.39 | 3.59 | SE2 | 3.77 | 4.57 | SE1 | 5.26 | 24 |
| J1545+5047 | 607 | 0.07 | 239 | NW1 | 7.97 | 100 | Total | 13 | 29 |  |  |  | SE1 | 8.21 | 113 |
|  | 1400 | 0.17 | 79 | NW1 | 4.23 | 30 | NW2 | 6.77 | 7.71 | SE2 | 4.02 | 4.81 | SE1 | 4.37 | 37 |
| J1605+0711† | 607 | 0.06 | 480 | N1 | 4.97 | 183 | N2 | 30 | 92 | S2 | 11 | 77 | S1 | 3.86 | 142 |
|  | 1400 | 0.18 | 159 | N1 | 5.19 | 64 | N2 | 15 | 33 | S2 | 5.17 | 18 | S1 | 1.85 | 45 |
| J1627+2906 | 607 | 0.10 | 278 | N1 | 33 | 138 | N2 | 11 | 16 | S2 | 11 | 14 | S1 | 16 | 118 |
|  | 1400 | 0.15 | 92 | N1 | 14 | 47 | N2 | 3.56 | 4.5 | S2 | 4.31 | 5.49 | S1 | 6.63 | 40 |
| J1649+4133 | 607 | 0.13 | 84 | NE1 | 17 | 45 | Total | 5.05 | 12 |  |  |  | SW1 | 8.01 | 32 |
|  | 1400 | 0.14 | 23 | NE1 | 5.41 | 12 | NE2 | 1.64 | 1.28 | SW2 | 1.41 | 1.18 | SW2 | 3.57 | 8.13 |
| J1705+3940 | 607 | 0.11 | 276 | NE1 | 17 | 92 | NE2 | 18 | 24 | SW2 | 40 | 54 | SW2 | 8.29 | 106 |
|  | 1400 | 0.14 | 76 | NE1 | 6.40 | 22 | NE2 | 6.88 | 8.50 | SW2 | 17 | 22 | SW2 | 2.96 | 28 |

Column 1: Name of the source; Column 2: Frequencies of observations; Column 3: The rms noise of the maps; Column 4: Integrated flux density from GMRT and FIRST maps; Columns 5,8,11,14: component designation, where W, E, S and N denote west, east, south and north components respectively. Numbers 1 and 2 indicate components formed by first and second epochs of activity. Total represents both outer or inner components. Column 6 and 7, 9 and 10, 12 and 13, 15 and 16: the peak flux densities and total flux densities measured from the total intensity maps for each component of the sources in units of mJy/beam and mJy respectively. These values have been estimated from the full-resolution GMRT images, where the beam sizes are listed in Table 1.

† Sources with lobe sizes much larger than $\sim 1'$ for which FIRST flux density estimates at 1400 MHz may be underestimated.

values of CI are consistent with 2.86 within the errors of measurement. Only 2 of the sources (J1326+1924 and J1328+2752) appear to have values larger than 2.86. For rest of the 5 sources in our sample CI values were not calculated as SDSS/Pan-STARRS did not provide reliable values of Petrosian radius. Thus 12 of the 16 sources with estimated values of CI have values either less than or consistent with 2.86. We carried out a visual examination of the images of all the above mentioned galaxy-fields obtained from SDSS catalogue and found that they are either point-like or very-faint objects (except the sources J1326+1924 and J1328+2752). So, we cannot conclude about their host morphologies on the basis of the data available to us. However, the tendency for radio sources showing recurrent activity to have lower values of CI is consistent with the trend reported by Kuźmicz et al. (2017), who suggested that this might be due to radio sources showing recurrent activity occuring more often in galaxies with disturbed/amorphous structures.

### 3.1 Discussion on individual sources

**J0746+4526:** This is a red quasar with evidence of recurrent activity (Nandi et al. 2014). Most double-double radio sources are associated with galaxies but it is also possible for a quasar to exhibit episodic activity. We present its GMRT 607-MHz image in Figure 1 (1st row, left panel) adapted from Nandi et al. (2014). One-dimensional cuts through the 607 MHz and 1400 MHz maps taken along the major axis of the source are presented below it. For both frequencies we can see two sharp brightness peaks for inner double and asymmetric brightness distribution for the outer double. The spectral index of the outer emission is steeper than the inner components.

**J0804+5809:** Our GMRT image at 607 MHz (Figure 1, 1st row, middle panel) shows the entire structure of J0804+5809. Here the young edge-brightened inner structure is completely embedded within diffuse relic emission which is not characterized by classical FRII structure. From the Half Million Quasars (HMQ) Catalogue (Flesch 2015) we have identified the host of this target as a candidate quasar, consistent with its WISE colours. Only two quasars (J0746+4526 and J0935+0204) with multiple jet activity are known to date (Nandi et al. 2014; Jamrozy et al. 2009). J0804+5809 is the third clear example of a candidate quasar which shows recurrent jet activity as well. The object also appears stellar in the Pan-STARRS image; however a spectroscopic confirmation is required. The outer structure is extended up to 610 kpc while the inner double has a separation of 94 kpc. The one dimensional cuts of the 607-MHz and 1400-MHz maps indicate high brightness distribution for both inner hotspots whereas the outer relic emission shows a low brightness distribution.

**J0855+4204:** Both GMRT (Figure 1, 1st row, right panel) and FIRST images of this source resolve the central component into a compact double with a separation of 36 kpc whereas the outer continuous bridge of emission extends up to 609 kpc. The brightness immediately falls from both sides of the inner hotspots, and a symmetric low brightness for both outer lobes is observed for this source.

**J0910+0345:** The 607-MHz image (Figure 1, 3rd row, left panel) of this source shows a barely resolved compact double lobed structure lying inside faint radio emission. A high resolution observation is required to reveal its detailed structure. The outer large-scale structure has a linear size of 396 kpc whereas the inner component extends only 36 kpc. In this case we also noticed a sharp fall of brightness distribution beyond the inner double and relatively symmetric low brightness distribution for both components of the outer double.

**J1039+0536** Our GMRT map (Figure 1, 3rd row, middle panel) shows the lobes of the inner double to be well separated by $\sim$ 29 kpc and lying on the opposite sides of the parent optical galaxy. The outer double extends up to 220 kpc. One-dimensional brightness distributions show that the peak brightness of the inner and outer doubles are comparable.

**J1103+0636** The GMRT image shows that the outer diffuse lobes have a linear extent of 887 kpc, without any hotspots. In Figure 1 (3rd row, right panel) the optical host appears in between the inner double with prominent hotspots. The inner hotspots have a projected separation of 71 kpc. Both the outer double as well as the inner double are reasonably symmetric in flux density.

**J1208+0821:** The GMRT image (Figure 2, 1st row, left panel) shows that both the outer lobes are edge brightened, with their peak brightnesses higher than those of the inner lobes. Higher resolution observations are required to examine their detailed structure. The inner double with bright hotspots is symmetric about the optical host and the hotspots are well separated from each other. The projected linear sizes of the inner and outer doubles of this source are $\sim$119 and $\sim$791 kpc respectively.

**J1238+1602:** Figure 2 (3rd row, left panel) shows the GMRT image of J1238+1602. Higher-resolution observations are required to clarify whether the outer lobes have hot-spots. The inner double is prominent with the northwestern component showing an edge-brightened structure. The source has no redshift information. The angular size of the outer emission is 120″ while that of the inner double is 44″. The north west and south east jet emission are not aligned with each other for each episode.

**J1240+2122:** For this source the 607-MHz image (Figure 2, 3rd row, middle panel) shows bright hotspots at the ends and a central component. The separation between the outer hotspots is 821 kpc. Any signature of back flow or bridge of emission is not detected in this source. A prominent gap in emission between the central double and the outer double is seen in both the GMRT and FIRST maps. We note that the inner double (linear size 117 kpc) is not collinear with the outer hotspots. Its inner component show a steeper spectral index than the outer components. Both the outer and inner doubles are highly asymmetric in intensity.

**1326+1924:** The FIRST image shows the inner 37 kpc double-lobed structure The GMRT image (Figure 2, 1st row, middle panel) with a resolution 7.73″×4.24″ does not reveal the detailed structure of the inner double. For the outer double (252 kpc), GMRT image shows a diffuse complex structure similar to an S-shaped source (Capetti et al. 2002) without any prominent hotspots. A merging black hole system can give rise to jet precession and such radio morphology (Merritt & Ekers 2002; Rubinur et al. 2017).

**J1328+2752:** J1328+2752 is a significantly misaligned DDRG. Figure 2 (3rd row, right panel; adopted from Nandi et al. 2017) shows the inner double to be completely misaligned with the outer double by an angle of $\sim$30°. An active hot spot still exists in the outer southern lobe while the outer northern lobe from an earlier cycle shows a curved structure. The linear extent of the outer and inner doubles are 413 kpc and 96 kpc respectively. This source is also hosted by a giant elliptical with double-peaked emission lines. Both radio morphology and optical emission lines indicate the source is plausibly associated with a merging massive black hole binary (Nandi et al. 2017).

**J1344−0030:** The GMRT image of this source is shown in Figure 2 (1st row, right panel). The faint outer emission is seen clearly in the low-frequency image and its extent is 747 kpc. The inner double (98 kpc) has prominent hotspots. A bridge of emission connects the outer and inner structure.

**J1407+5132:** In Figure 3 (1st row, left panel) we present 607-MHz image of this target. A prominent core emission has been noticed in the available 4860-MHz VLA image whereas FIRST and GMRT images do not show such emission. The outer and inner doubles are extended upto 866 kpc and 91 kpc respectively. One dimensional brightness distributions for GMRT and FIRST data show the peak brightness of the outer doubles to be larger than the inner ones, although there is no evidence of clear hotspots. For this source the spectral index of inner component is steeper than the outer one.

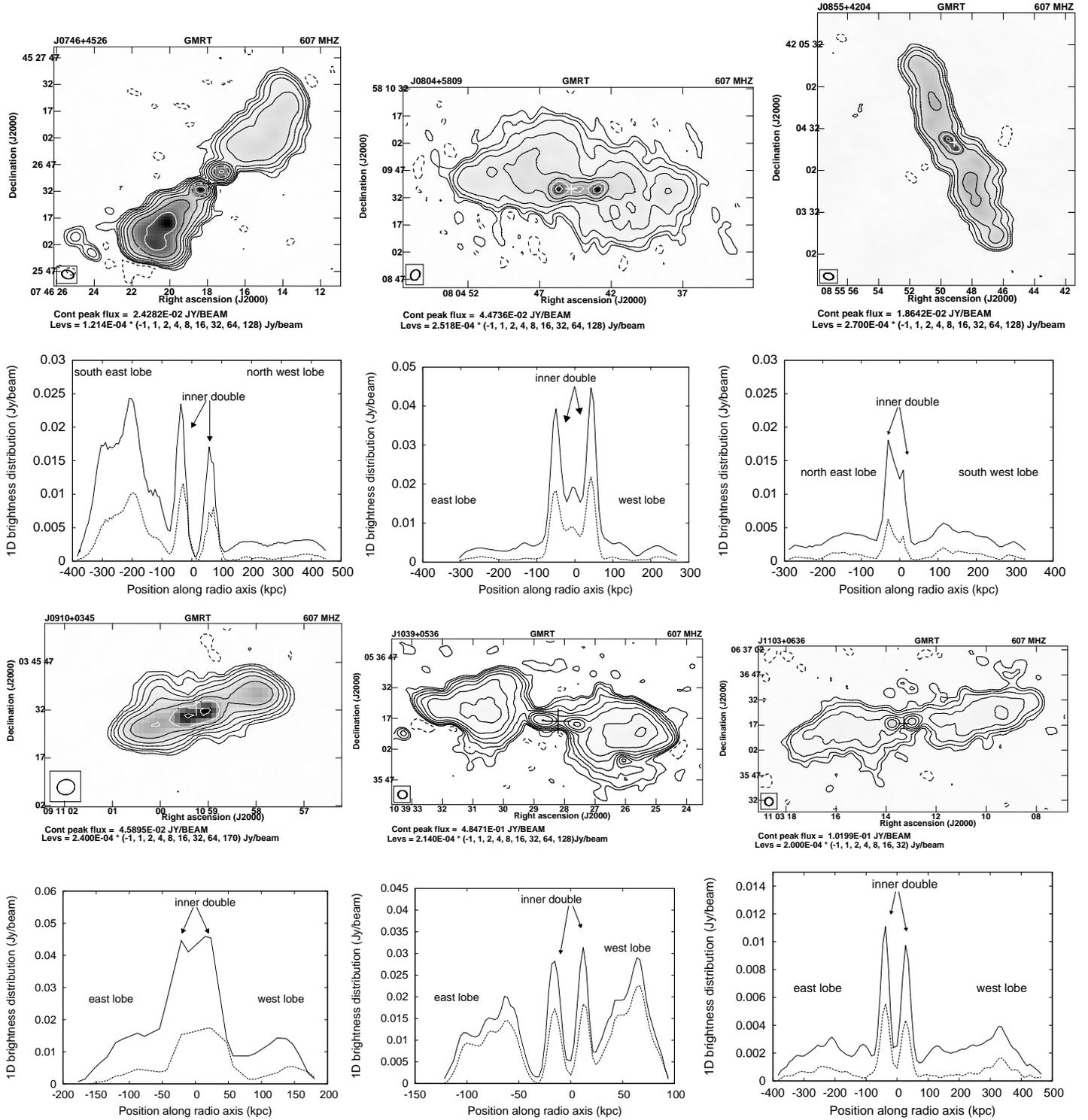

**Figure 1.** 1st row: GMRT 607-MHz images of J0746+4526, J0804+5809 and J0855+4204. 2nd row: One-dimensional brightness distributions at 607 MHz (solid line) and 1400 MHz (dashed line) respectively for each of the above mentioned DDRGs. The cuts taken along the radio axes and passing through the central hosts of J0746+4526 J0804+5809 and J0855+4204 are at PAs of 135°, 88° and 28° respectively. 3rd row: GMRT 607-MHz images of J0910+0345, J1039+0536 and J1103+0636. 4th row: One-dimensional brightness distributions at 607 MHz (solid line) and 1400 MHz (dashed line) respectively for each the DDRGs mentioned in 3rd row. The cuts taken along the radio axes and passing through the central hosts of J0746+4526 J0804+5809 and J0855+4204 are at PAs of 110°, 261° and 96° respectively. In all GMRT images the + sign denotes the position of the optical object.

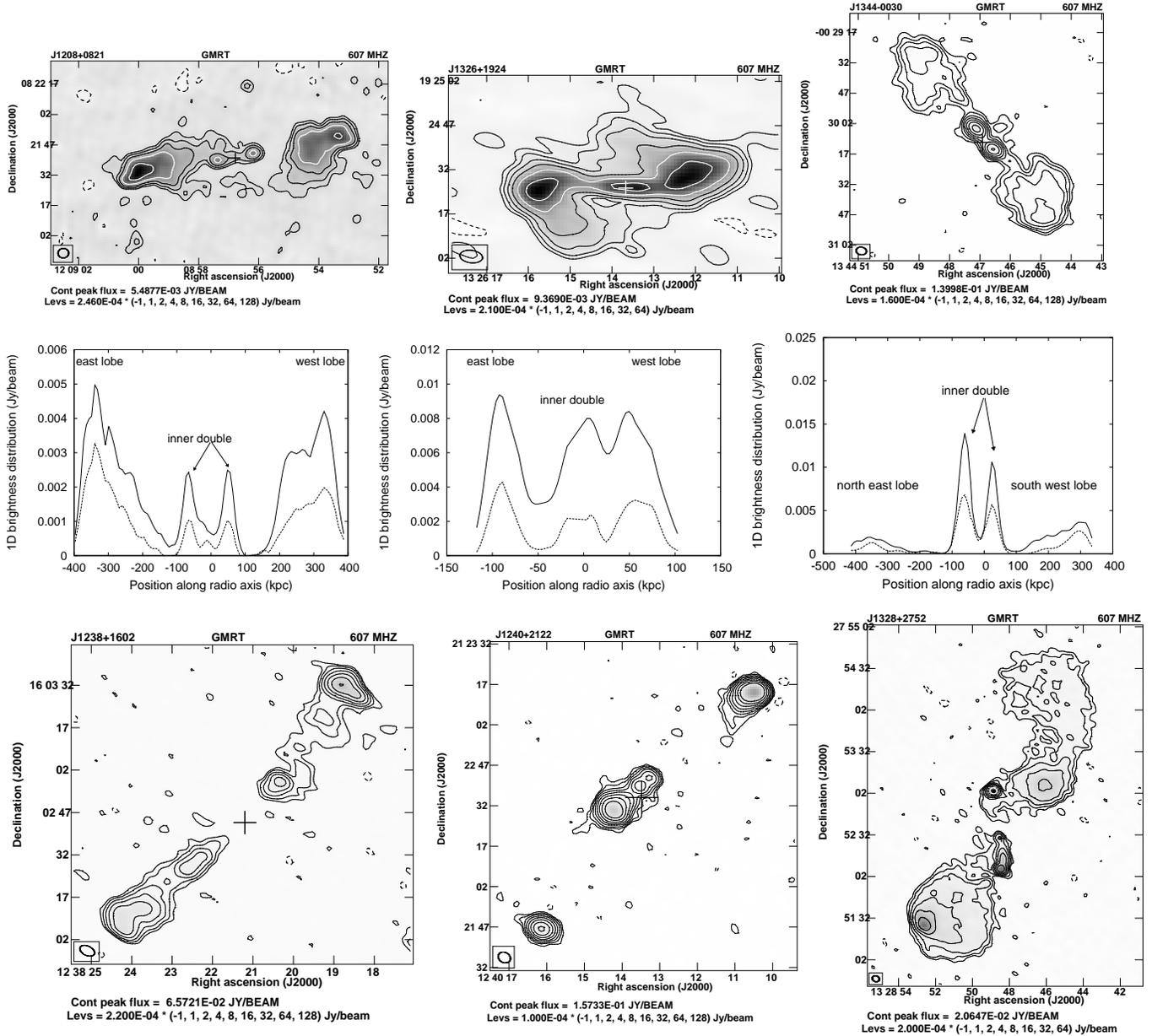

**Figure 2.** 1st row: GMRT 607-MHz images of J1208+0821, J1326+1924 and J1344−0030. 2nd row: One-dimensional brightness distributions at 607 MHz (solid line) and 1400 MHz (dashed line) respectively for each of the above mentioned DDRGs. The cuts taken along the radio axes and passing through the central hosts of J1208+0821, J1326+1924 and J1344−0030 are at PAs of 102°, 92° and 220° respectively. 3rd row: GMRT 607-MHz images of J1238+1602, J1240+2122 and J1328+2752. All these three sources are misaligned and their one-dimensional brightness distributions are not possible to plot. In all GMRT images the + sign denotes the position of the optical object.

**J1500+5142:** We present the GMRT image of J1500+5142 in Figure 3 (1st row, middle panel). Well-resolved inner double has an asymmetric flux density distribution, with the southern inner lobe being much brighter. The outer lobes are highly diffuse and there is an emission gap in between north west outer and inner components. The projected linear size of outer and inner lobes are 608 kpc and 143 kpc respectively.

**J1521+5214:** The 610 MHz map of this target could not resolve the inner double well (see Figure 3, 1st row right panel), while this structure is clearly resolved in the FIRST image. The inner double is ∼92 kpc in extent, while the outer double extends to 559 kpc.

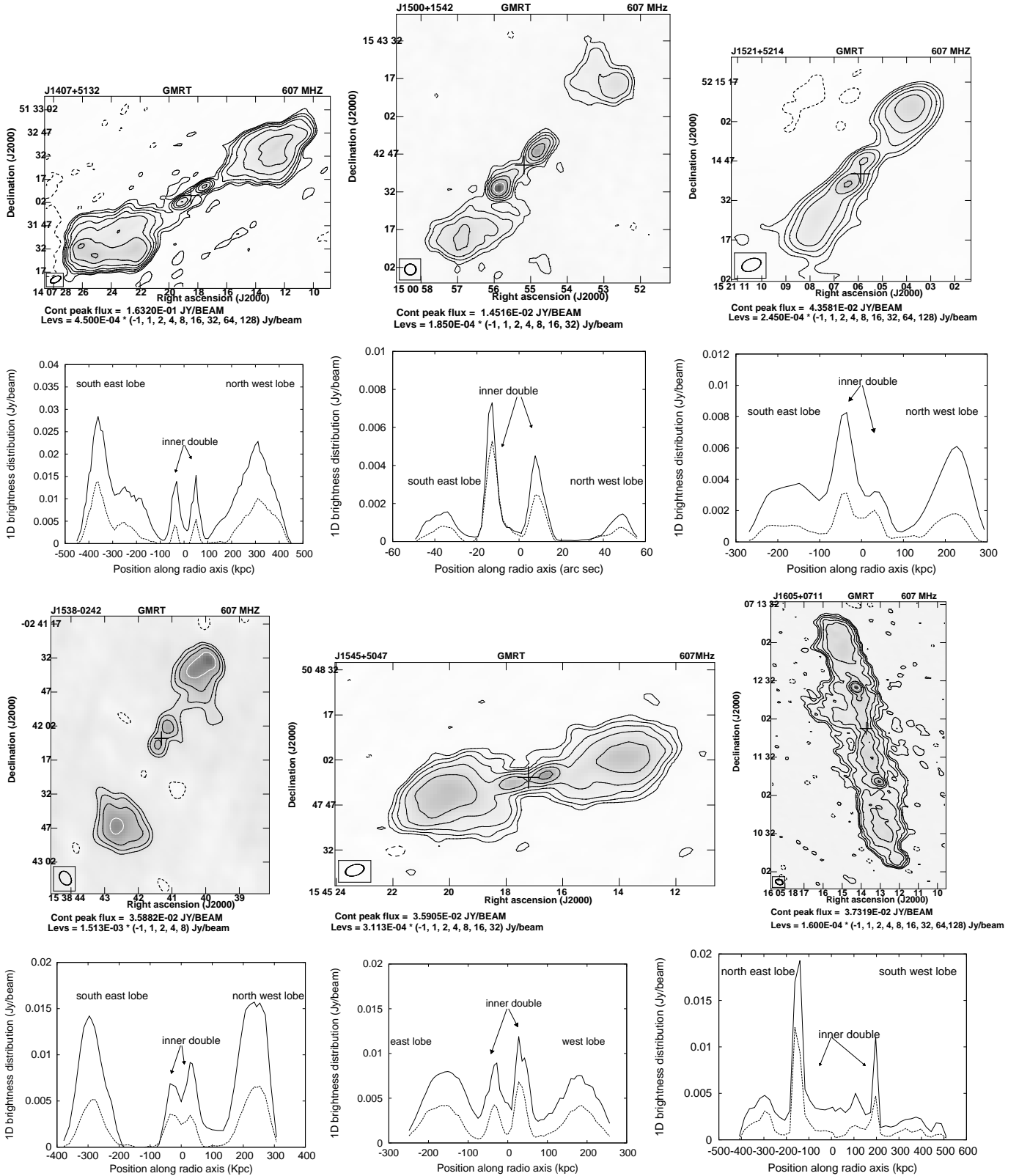

**Figure 3.** 1st row: GMRT 607-MHz images of J1407+5132, J1500+1542 and J1521+5114. 2nd row: One-dimensional brightness distributions at 607 MHz (solid line) and 1400 MHz (dashed line) respectively for each of the above mentioned DDRGs. The cuts taken along the radio axes and passing through the central hosts of J1407+5132, J1500+1542 and J1521+5114 are at PAs of 295°, 314° and 320° respectively. 3rd row: GMRT 607-MHz images of J1538−0242, J1545+5047 and J1605+0711. 4th row: One-dimensional brightness distributions at 607 MHz (solid line) and 1400 MHz (dashed line) respectively for each the DDRGs mentioned in 3rd row. The cuts taken along the radio axes and passing through the central hosts of J1538−0242, J1545+5047 and J1605+0711 are at PAs of 152°, 286° and 13° respectively. In all GMRT images the + sign denotes the position of the optical object.

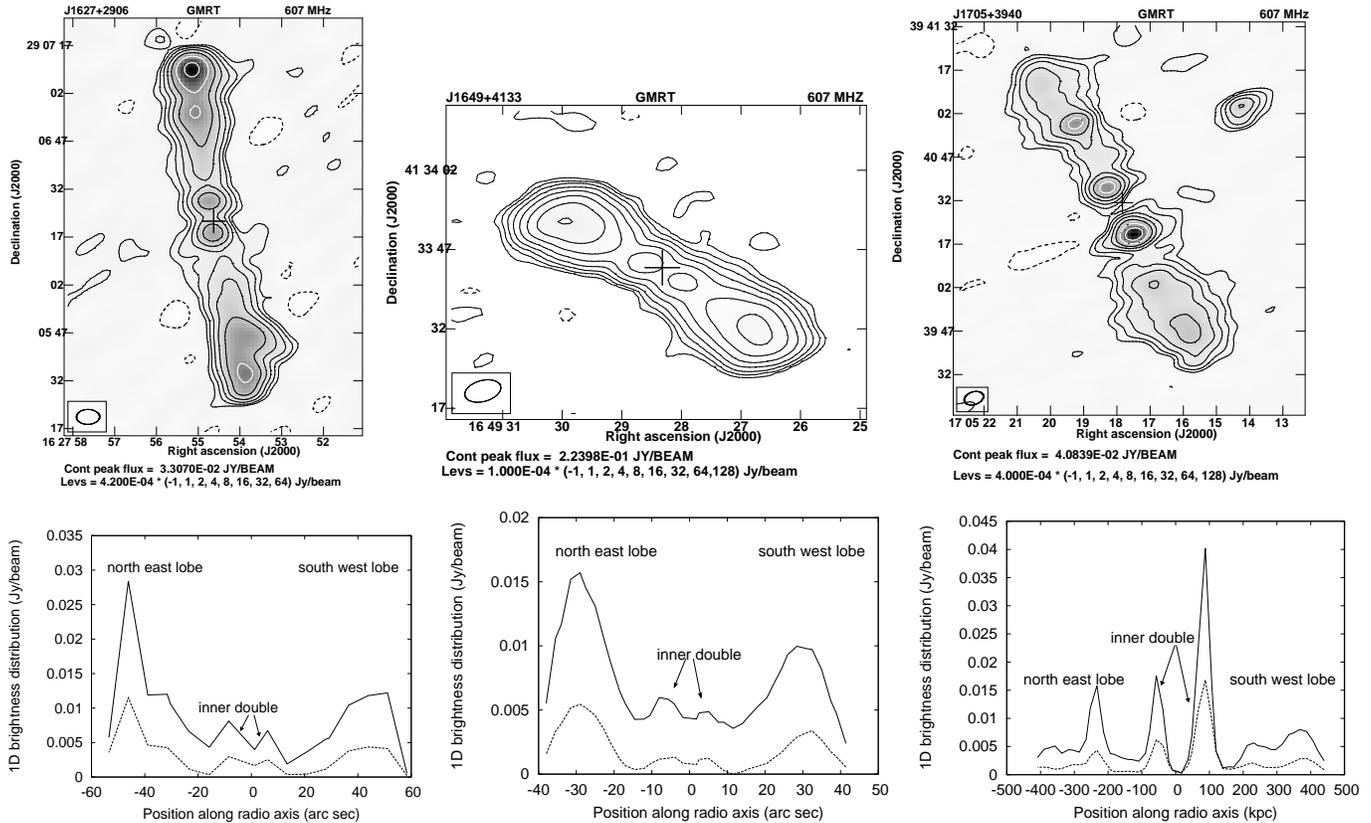

**Figure 4.** 1st row: GMRT 607-MHz images of J1627+2906, J1649+4133 and J1705+3940. 2nd row: One-dimensional brightness distributions at 607 MHz (solid line) and 1400 MHz (dashed line) respectively for each of the above mentioned DDRGs. The cuts taken along the radio axes and passing through the central hosts of J1627+2906, J1649+4133 and J1705+3940 are at PAs of 10°, 242° and 32° respectively. In all GMRT images the + sign denotes the position of the optical object.

**J1538−0242:** In Figure 3 (3rd row, left panel) the 610 MHz map of J1538−0242 has been shown. The outer and inner doubles have linear sizes of 658 and 62 kpc respectively. Here the peak brightness of the outer lobes is higher than the inner ones.

**J1545+5047:** Outer lobes of this target shows a symmetric morphology extending over 519 kpc. The inner double is surrounded by diffuse emission at 610 MHz (Figure 3, 3rd row middle panel). The FIRST map clearly resolves the inner double and its overall linear size is 71 kpc.

**J1605+0711:** In Figure 3 (3rd row, right panel) we present GMRT image of J1605+0711. In our sample this is the largest source where restarted jets clearly propagate through the cocoon of the previous active phase. One dimensional flux density cuts show compact hotspot peaks of the inner double separated by 349 kpc embedded within relict cocoon of 917 kpc. The morphology of this source is very similar to the well known remarkable DDRG J1548-3216 (Machalski et al. 2010).

**J1627+2906:** Both the FIRST and GMRT images (Fig. 4, 1st row left panel), show that the north-east outer lobe has a hotspot, while this is not clear for the south-western outer lobe. The outer emission extends 802 kpc, while the inner double is 88 kpc in extent with lobes of similar peak brightness.

**J1649+4133:** The inner double is embedded in the extended emission (Fig. 4, 1st row, middle panel). The outer relict emission extends about 52.8″ whereas inner component has an extension 8.4″. Redshift information is not available. The spectral index of the inner double is slightly steeper than the outer emission. This may be due to the mixing of emission from two epochs.

**J1705+3940:** Fig. 4 (1st row right panel) shows the GMRT image of J1705+3940. The northern outer double has a compact feature in the middle of the lobe which is of similar brightness to the northern inner double. There is also a weaker peak in the southern outer lobe. Higher-resolution observations are required to explore whether there might have been more than two cycles of activity. The projected linear size for inner double is 146 kpc while that of the outer emission is 865 kpc.

## 4 AGE OF THE SOURCES

In order to understand the timescales of episodic activity, it is crucial to estimate the ages of charged particles in different epochs. The spectral shape of radio sources helps to constrain the age of the electron population. We have compiled the available total flux densities at different frequencies for each of the sources to examine any evidence of spectral ageing since we do not have adequate data for the inner and outer doubles separately. Through the inspection of integrated spec-

**Table 3.** Comparison of spectral indices between outer and inner components

| Name | $\alpha_{607}^{1400}$ Outer | $\alpha_{607}^{1400}$ Inner | $l_{out}$ kpc | $l_{inn}$ kpc | CI |
|---|---|---|---|---|---|
| (1) | (2) | (3) | (4) | (5) | (6) |
| J0746+4526 | 1.27±0.10 | 0.90±0.10 | 887 | 101 | 2.89±0.84 |
| J0804+5809 | 1.88±0.10 | 0.95±0.10 | 610 | 94 | 2.75±0.01 |
| J0855+4204† | 1.26 | 0.86±0.10 | 609 | 36 | 2.74±0.22 |
| J0910+0345 | 1.57±0.10 | 0.88±0.10 | 396 | 36 | |
| J1039+0536 | 0.97±0.10 | 0.90±0.10 | 220 | 29 | 2.89±0.53 |
| J1103+0636 | 1.64±0.10 | 1.01±0.10 | 887 | 71 | ⩽3.35 |
| J1208+0821 | 1.10±0.10 | 1.04±0.10 | 791 | 119 | 2.03±0.66 |
| J1238+1602 | 0.84±0.10 | 0.76±0.10 | | | |
| J1240+2122 | 0.68±0.10 | 1.00±0.10 | 821 | 117 | ⩽1.78 |
| J1326+1924 | 1.30±0.10 | 1.25±0.10 | 252 | 37 | 3.15±0.24 |
| J1328+2752† | 1.23 | 0.90±0.10 | 413 | 96 | 3.25±0.15 |
| J1344−0030 | 1.47±0.10 | 0.86±0.10 | 747 | 98 | ⩽2.76 |
| J1407+5132† | 1.30 | 1.61±0.10 | 866 | 91 | 3.01±0.42 |
| J1500+1542 | 1.26±0.10 | 0.87±0.10 | 608 | 143 | |
| J1521+5214 | 1.33±0.10 | 0.85±0.10 | 559 | 92 | |
| J1538−0242 | 1.49±0.10 | 1.20±0.10 | 658 | 62 | 1.57±1.17 |
| J1545+5047 | 1.23±0.10 | 0.93±0.10 | 519 | 71 | ⩽2.99 |
| J1605+0711† | 1.41 | 1.36±0.10 | 917 | 349 | 2.82±0.31 |
| J1627+2906 | 1.21±0.10 | 1.15±0.10 | 802 | 88 | ⩽1.88 |
| J1649+4133 | 1.50±0.10 | 1.53±0.10 | | | |
| J1705+3940 | 1.49±0.10 | 1.12±0.10 | 865 | 146 | ⩽2.09 |

Column 1: name of the source; Columns 2 and 3: integrated spectral indices between 607 and 1400 MHz of the outer and inner doubles respectively; Columns 4 and 5: projected linear sizes of the outer and inner doubles respectively; Column 6: Concentration index.
† For these sources the spectral indices of outer doubles represent the upper limits.

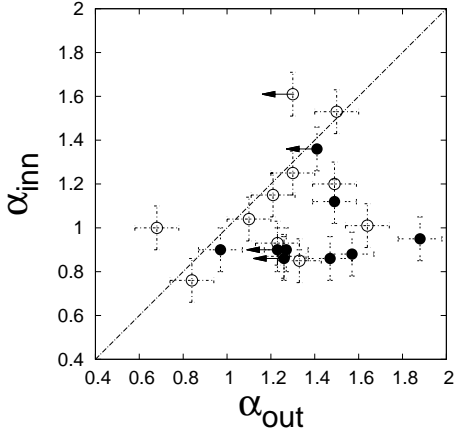

**Figure 5.** Spectral index variation of new sample of DDRGs. The open circles represent the sources which have flux density ≲5 mJy for inner components. While the filled circles represent the sources which have flux density ≳5 mJy for inner components. The dashed line represents no change in spectral index for outer and inner components. The upper limits to spectral indices indicate sources with components larger than ∼1′, where the FIRST 1400 MHz flux densities may be underestimate.

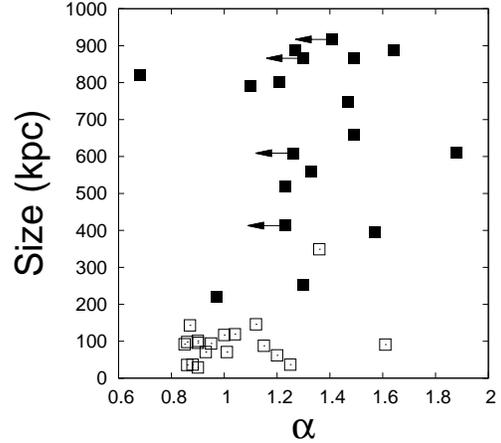

**Figure 6.** Sizes of the inner and outer doubles vs spectral index for this sample of DDRGs. Open boxes represent inner doubles while filled boxes represent outer doubles. The angular size of the outer lobes is above ∼1′ for 4 sources. The 1400 MHz flux densities of these sources may have been underestimated due to their large size. The spectral indices of their outer doubles represent the upper limits (solid arrow).

tra we noted marginal evidence of steepening for the following sources: J0746+4526, J0804+5809, J0855+4204, J1039+0536, J1326+1924, J1328+2752, J1545+5047 and J1627+2906. For these 8 sources total flux density at 607 MHz is dominated by the outer lobes as well. Their flux densities with respective references are listed in Table 4. The total flux densities at 607 MHz are taken from Table 2. We used NVSS (NRAO/VLA Sky Survey; Condon et al. 1998) measurement of the total flux densities at 1400 MHz.

### 4.1 Spectral ageing analysis

Spectral ageing studies have been reported for double-double radio sources by several authors (Jamrozy et al. 2008; Nandi et al. 2010; Konar et al. 2012). The spectral age of radio sources is determined using the classical synchrotron theory which describes the time evolution of the emission spectrum from particles with an initial power-law energy distribution characterized by an injection spectral index that can be estimated from low-frequency data. After sufficient amount of time has elapsed, the radio spectrum steepens above a break frequency due to radiative cooling of high energy particles. The break frequency above which the radio spectrum steepens, $\nu_{\rm br}$, is related to the radiative synchrotron age, $\tau_{\rm syn}$, and the magnetic field strength, $B$, through the following relation.

$$\tau_{\rm syn}[{\rm Myr}] = 50.3 \frac{B^{1/2}}{B^2 + B_{\rm iC}^2}[\nu_{\rm br}(1+z)]^{-1/2}, \quad (2)$$

where $B_{\rm iC}=0.318(1+z)^2$ is the magnetic field strength equivalent to the inverse-Compton microwave background radiation. $B$ and $B_{\rm iC}$ are in units of nT and $\nu_{\rm br}$ is in GHz.

For the above-mentioned sources we tried to estimate break frequency and hence radiative ages for the entire structure. For these sources we fitted the spectra with Jaffe & Perola (1973, JP) model using SYNAGE package (Murgia et al. 1999). For each fit (Fig. 7) we treat injection spectral index ($\alpha_{inj}$) as a free parameter as well as a fixed parameter. We found that fits are better if we keep ($\alpha_{inj}$) as a free parameter. The SYNAGE fits gave high break frequencies with large uncertainties, possibly due to the marginal evidence of any steepening. Since the integrated flux densities of these sources are available up to $\sim 5$ GHz, we made a very conservative estimate of an upper limit to its age by considering the lower limit to the break frequency to be $>5$ GHz. To estimate the magnetic fields we follow similar method to that of Nandi et al. (2010) and Konar et al. (2012). The measured field values and radiative upper age limits are given in table 5. For these sources the flux density from inner components are $\lesssim 15\%$ of integrated flux densities at 607 MHz for all but one source (see table 5). So, the main flux density contribution comes from the outer lobes for each source. Our measurements represent a first order estimate of the spectral age of the particle population for the outer emission.

### 4.2 Time-scale of jet interruption

Presence of a hotspot in an outer lobe in a DDRG is useful to estimate the time-scale of jet interruption (Jamrozy et al. 2009; Joshi et al. 2011; Konar et al. 2012). The hotspots in the inner lobes represent the termination points of current jet flow while the hotspot in the outer lobes indicate that these are still receiving material from previous jet activity. For the sake of simplicity we assume sources are symmetrical. There are hotspots detected in the outer lobes for J1240+2122, J1328+2752 and J1627+2906. Assuming an inclination angle of $45°$ and a jet velocity of 0.5c (Jamrozy et al. 2009) we estimate jet switch-off time for these sources. This result will be affected by light-travel time effects because of the orientation of the source. We follow Konar et al. (2012) to estimate the time-scale of jet interruption. We assume previous jet activity turns off on both sides at same time. If the previous jet stopped $t_{jet}$ time ago then that jet material will flow through old jet channel with velocity 0.5c. The condition for the absence of hotspot in outer lobe approaching us, is $0.5ct_{jet}>L$; where L is the distance from the core to the hotspot. The hotspot which is still present in the other side of the core must be on the receding side. So the light travel time $(D/c)$ is longer for this hotspot. D is the distance travelled by the photon from the receding hotspot. Hence the condition for presence of hotspot far side of the source is $0.5c(t_{jet}-D/c)<L$. Here D= 2Lcos$\theta$ and the observed source length $L_{obs} = 2$Lsin$\theta$ for angle to the line of sight $\theta$. For sources J1328+2752 and J1627+2906 which have only one visible hot-spot in the outer lobes, the maximum time difference between the hotspots are $\sim 1.34$ and $\sim 2.6$ Myr respectively. On the other hand, J1240+2122 appears to have hot-spots on both sides of the earlier cycle of activity. For this source the maximum time difference between the hotspots is 2.67 Myr. Considering the above, for J1328+2752, we find that the jet switch-off time, measured with respect to the hotspot on the receding side, must be more than $\sim 1.9$ Myr and less than $\sim 3.25$ Myr ago. Similarly for the source J1627+2906 jet switch off time is more than $\sim 3.69$ Myr ago and less than $\sim 6.3$ Myr ago. For J1240+2122, both outer lobes are still fuelled by the earlier epoch of jet activity. For a similar jet velocity and inclination angle, the jet switch off time is less than about 3.8 Myr.

### 5 SUMMARY AND CONCLUDING REMARKS

We summarize briefly the main conclusions of the paper.

(i) Here we have presented the radio continuum 607 MHz images of 21 candidate double-double radio galaxies identified from the FIRST survey. For most of the sources, these new low-frequency observations reveal more diffuse emission from the outer fossil lobes than the FIRST survey.

(ii) Using both 607-MHz and the FIRST images we estimated the spectral index for the inner and outer lobes of each source. Our study shows that none of the inner compact components is a core with a flat radio spectrum. This confirms the episodic nature of these radio sources. For the vast majority of sources, the spectral index of the outer lobes is significantly steeper (Figs. 5 and 6), as expected for DDRGs. However, for sources J1240+2122 and 1407+5132 outer emission appears to have a flatter spectral index than the inner double. These objects lie significantly above the dashed line shown in Fig.(5). For J1240+2122 we note the presence of compact hotspots without any diffuse emission in both of the outer lobes. It is important to confirm this from observations over a larger frequency range as it might indicate a difference in the electron injection spectral indices in the two epochs.

(iii) The object J0746+4526, is a double double radio quasar and it has already been reported in Nandi et al. (2014). We also identified J0804+5809 as a candidate quasar with recurrent activity.

(iv) We find that the two epochs of jet activity are not collinear for the sources J1238+1602, J1240+2122 and J1328+2752. In the case of J1326+1924 GMRT could not resolve well the inner structure and the outer diffuse emission is quite similar to S shaped radio galaxies, suggesting precession of the jet axis between the two epochs.

(v) For eight sources which showed marginal evidence of spectral steepening in their integrated spectra from all

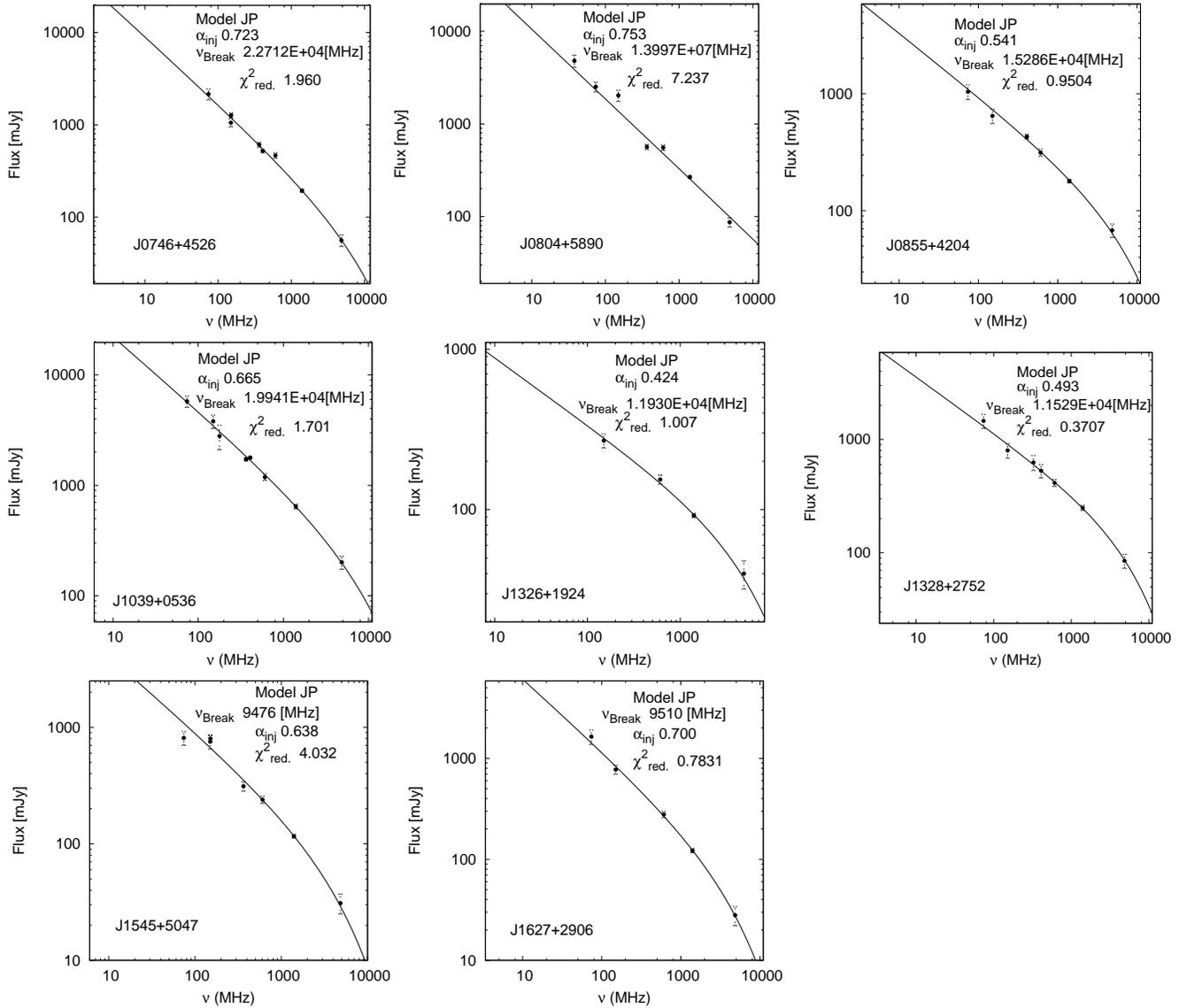

**Figure 7.** Spectra for the 8 sources and the fits using the JP model. The injection spectral indices and break frequencies derived from the best fits are shown for each source.

available flux density measurements, we estimated an upper limit to the corresponding particle ages by assuming a lower limit to the break frequency of 5 GHz. The obtained age limits vary from $\sim$ 11 Myr to 52 Myr. This is relatively lower lifetime value in comparison to remnant radio sources (Jamrozy et al. 2004; Murgia et al. 2011; Brienza et al. 2016) as well as large sized DDRGs (Saikia & Jamrozy 2009).

(vi) The sources J1240+2122, J1328+2752, J1627+2906 show hotspots in the edges of the outer lobes in GMRT images. The time-scale of jet interruption for these sources have been estimated to be typically smaller than a few Myr, which are smaller than the large-sized DDRGs with diffuse outer lobes (e.g., Konar et al. 2012, and references therein).

(vii) For eight sources we note the peak brightness of outer double is higher than the inner young double. The enhancements in surface brightness detected at the edges of the outer lobes may may indicate that the central AGN has stopped and restarted within a short amount of time. We need high frequency sensitive observations in order to detect these remnant 'hotspots' created by earlier jet activity. We plan to make multi frequency high resolution radio observations of these sources to estimate the spectral age for inner and outer lobes separately. These data will also be necessary to place better constraints on the jet intermittent time scale of these sources.

## ACKNOWLEDGMENTS

We acknowledge the anonymous referee for his/her valuable comments that have improved the paper significantly. SN is funded by Wenner-Gren foundation (Stockholm, Sweden) to conduct her research projects. SN is thankful to DST, Government of India and Belgian Federal Science Policy (BEL-SPO) for financial assistance during which period the observations and other works were carried out. We thank the

**Table 4.** Integrated flux densities

| Object name (1) | Obs. Freq. (MHz) (2) | S (mJy) (3) | error (mJy) (4) | Ref used (5) |
|---|---|---|---|---|
| J0746+4526 | 74 | 2150 | 290 | (1) |
| | 150 | 1052 | 105 | (2) |
| | 151 | 1270 | 75 | (3) |
| | 365 | 609 | 40 | (4) |
| | 408 | 520 | 20 | (5) |
| | 607 | 466 | 32 | (6) |
| | 1400 | 193 | 7 | (7) |
| | 4850 | 56 | 8 | (8) |
| J0804+5809 | 38 | 4800 | 720 | (9) |
| | 74 | 2520 | 310 | (1) |
| | 150 | 2032 | 203 | (2) |
| | 365 | 566 | 36 | (4) |
| | 607 | 556 | 39 | (6) |
| | 1400 | 267 | 9 | (7) |
| | 4850 | 87 | 10 | (8) |
| J0855+4204 | 74 | 1040 | 150 | (1) |
| | 150 | 646 | 64 | (2) |
| | 408 | 430 | 20 | (5) |
| | 607 | 314 | 22 | (6) |
| | 1400 | 179 | 6 | (7) |
| | 4850 | 68 | 9 | (8) |
| J1039+0536 | 74 | 5760 | 670 | (1) |
| | 150 | 3808 | 381 | (2) |
| | 178 | 2800 | 700 | (10) |
| | 365 | 1720 | 56 | (4) |
| | 408 | 1780 | 50 | (11) |
| | 607 | 1191 | 83 | (6) |
| | 1400 | 642 | 32 | (7) |
| | 4850 | 201 | 28 | (8) |
| J1326+1924 | 150 | 269 | 27 | (2) |
| | 607 | 154 | 10 | (6) |
| | 1400 | 92 | 3.4 | (7) |
| | 4850 | 40 | 8 | (8) |
| J1328+2752 | 74 | 1457 | 208 | (12) |
| | 151 | 800 | 117 | (13) |
| | 325 | 627 | 94 | (14) |
| | 408 | 529 | 72 | (15) |
| | 607 | 414 | 29 | (6) |
| | 1400 | 249 | 12 | (7) |
| | 4850 | 85 | 12 | (8) |
| J1545+5047 | 74 | 810 | 110 | (1) |
| | 150 | 754 | 75 | (2) |
| | 151 | 800 | 50 | (16) |
| | 365 | 312 | 30 | (4) |
| | 607 | 239 | 17 | (6) |
| | 1400 | 116 | 4 | (7) |
| | 4850 | 31 | 6 | (8) |
| J1627+2906 | 74 | 1640 | 270 | (1) |
| | 150 | 774 | 77 | (2) |
| | 607 | 278 | 19.5 | (6) |
| | 1400 | 122 | 6 | (7) |
| | 4850 | 28 | 6 | (8) |

Column 1 gives the name of the object. Column 2 gives the frequency, Columns 3 and 4 give the total flux densities of the source and the error, Column 5 gives references - (1) Cohen et al. (2007) (2)TGSS, (3) Hales et al. (1993), (4) Douglas et al. (1996), (5) Ficarra et al. (1985), (6) our observation, (7) Condon et al. (1998),(8) Gregory & Condon (1991), (9) Hales et al. (1995), (10) Gower et al. (1967), (11) Large et al. (1981), (12) Lane et al. (2014), (13) Waldram et al. (1996), (14) GMRT archival data, (15) Colla et al. (1972), (16) Hales et al. (1988).

**Table 5.** Magnetic field and spectral age estimates

| Object name (1) | $S_{in}/S_I$ % (2) | B (nT) (3) | $\tau_{\rm syn}$ (Myr) (4) |
|---|---|---|---|
| J0746+4526 | 12 | 0.30 | $\lesssim$15 |
| J0804+5809 | 31 | 0.23 | $\lesssim$28 |
| J0855+4204 | 15 | 0.28 | $\lesssim$34 |
| J1039+0536 | 9 | 0.42 | $\lesssim$44 |
| J1326+1924 | 11 | 0.23 | $\lesssim$40 |
| J1328+2752 | 13 | 0.18 | $\lesssim$52 |
| J1545+5047 | 12 | 0.41 | $\lesssim$20 |
| J1627+2906 | 11 | 0.56 | $\lesssim$11 |

Column 1: source name; Column 2: Percentage of the 607-MHz flux density originating from the inner lobes; Column 3: magnetic fields in nT; Column 4: upper limits to the synchrotron age in Myr

GMRT staff for technical support during the observations. GMRT is run by the National Centre for Radio Astrophysics of the Tata Institute of Fundamental Research. This research has made use of the NASA/IPAC Extragalactic Database (NED) which is operated by the Jet Propulsion Laboratory, California Institute of Technology, under contract with the National Aeronautics and Space Administration. Funding for the SDSS and SDSS-II has been provided by the Alfred P. Sloan Foundation, the Participating Institutions, the National Science Foundation, the U.S. Department of Energy, the National Aeronautics and Space Administration, the Japanese Monbukagakusho, the Max Planck Society, and the Higher Education Funding Council for England.